\documentclass[journal]{vgtc}                % final (journal style)
\ifpdf%                                % if we use pdflatex
  \pdfoutput=1\relax                   % create PDFs from pdfLaTeX
  \usepackage{graphicx}                % allow us to embed graphics files
  \DeclareGraphicsExtensions{.pdf,.png,.jpg,.jpeg} % for pdflatex we expect .pdf, .png, or .jpg files
\else%                                 % else we use pure latex
  \usepackage{graphicx}                % allow us to embed graphics files
  \DeclareGraphicsExtensions{.eps}     % for pure latex we expect eps files
\fi%

\graphicspath{{figures/}{pictures/}{images/}{./}} % where to search for the images

\usepackage{microtype}                 % use micro-typography (slightly more compact, better to read)
\PassOptionsToPackage{warn}{textcomp}  % to address font issues with \textrightarrow
\usepackage{textcomp}                  % use better special symbols
\usepackage{mathptmx}                  % use matching math font
\usepackage{times}                     % we use Times as the main font
\usepackage{cite}                      % needed to automatically sort the references
\usepackage{tabu}                      % only used for the table example
\usepackage{booktabs}                  % only used for the table example
\usepackage{float}

\usepackage{enumitem}
\onlineid{0}

\vgtccategory{Research}
\vgtcpapertype{application/design study}

\title{Exploring the Human Connectome Topology in Group Studies}

\author{Johnson J.G. Keiriz, Liang Zhan, Morris Chukhman, Olu Ajilore, Alex D. Leow, and Angus G. Forbes}
\authorfooter{
\item
 Johnson J.G. Keiriz, Morris Chukhman, Angus G. Forbes, Olu Ajilore, and Alex D. Leow  are with University of Illinois at Chicago. 
 E-mail: \{jgadel2, chukhman, aforbes\}@uic.edu; \{oajilore, aleow\}@psych.uic.edu
 \item
Liang Zhan is with University of Wisconsin--Stout. Email: zhan.liang@gmail.com
}

\shortauthortitle{Keiriz \MakeLowercase{\textit{et al.}}: Exploring the Human Connectome Topology in Group Studies}

\abstract{Visually comparing brain networks, or connectomes, is an essential task in the field of neuroscience. Especially relevant to the field of clinical neuroscience, group studies that examine differences between populations or changes over time within a population enable neuroscientists to reason about effective diagnoses and treatments for a range of neuropsychiatric disorders. In this paper, we specifically explore how visual analytics tools can be used to facilitate various clinical neuroscience tasks, in which observation and analysis of meaningful patterns in the connectome can support patient diagnosis and treatment. We conduct a survey of visualization tasks that enable clinical neuroscience activities, and further explore how existing connectome visualization tools support or fail to support these tasks. Based on our investigation of these tasks, we introduce a novel visualization tool, \textit{NeuroCave}, to support group studies analyses. We discuss how our design decisions (the use of immersive visualization, the use of hierarchical clustering and dimensionality reduction techniques, and the choice of visual encodings) are motivated by these tasks. We evaluate \textit{NeuroCave} through two use cases that illustrate the utility of interactive connectome visualization in clinical neuroscience contexts. In the first use case, we study sex differences using functional connectomes and discover hidden connectome patterns associated with well-known cognitive differences in spatial and verbal abilities. In the second use case, we show how the utility of visualizing the brain in different topological space coupled with clustering information can reveal the brain's intrinsic structure.
} % end of abstract


\CCScatlist{ % not used in journal version
 \CCScat{K.6.1}{Management of Computing and Information Systems}%
{Project and People Management}{Life Cycle};
 \CCScat{K.7.m}{The Computing Profession}{Miscellaneous}{Ethics}
}

\teaser{
  \centering
  \includegraphics[width=6in]{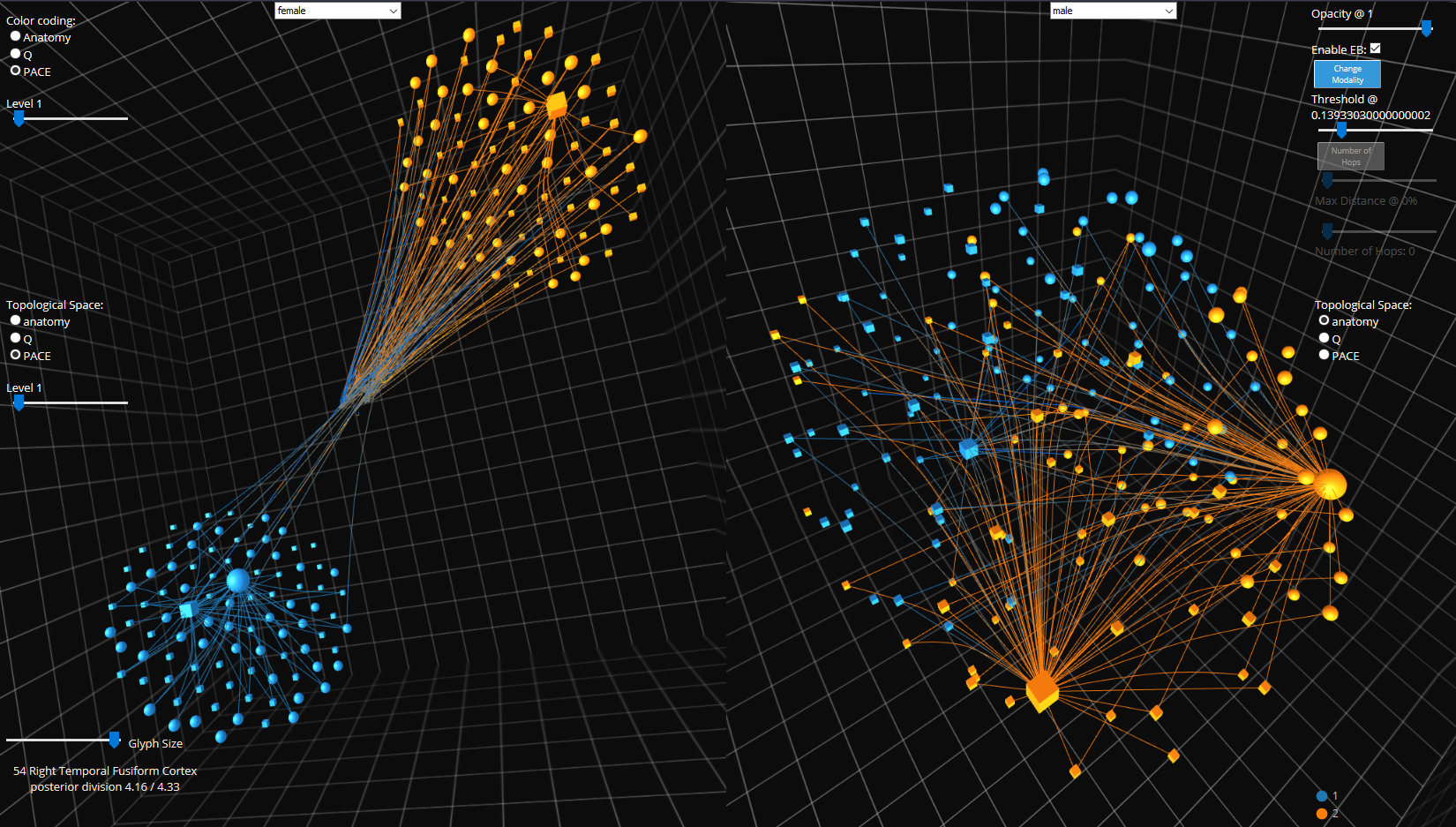}
  \caption{A screen capture of the \textit{NeuroCave} visualization tool showing the average resting state functional connectome of subjects between the age of 20 and 30 years old for females in the clustering space (left) versus males in the anatomical space (right). Our tool facilitates the simultaneous exploration of multiple connectome datasets in a variety of configurations, enabling researchers to make meaningful comparisons between them and to reason about their differences.
   }
	\label{fig:teaser}
}

\vgtcinsertpkg

\begin{document}

\firstsection{Introduction} \label{intro}

\maketitle

\begin{figure*}[htp]
 \centering
 \includegraphics[width=\textwidth]{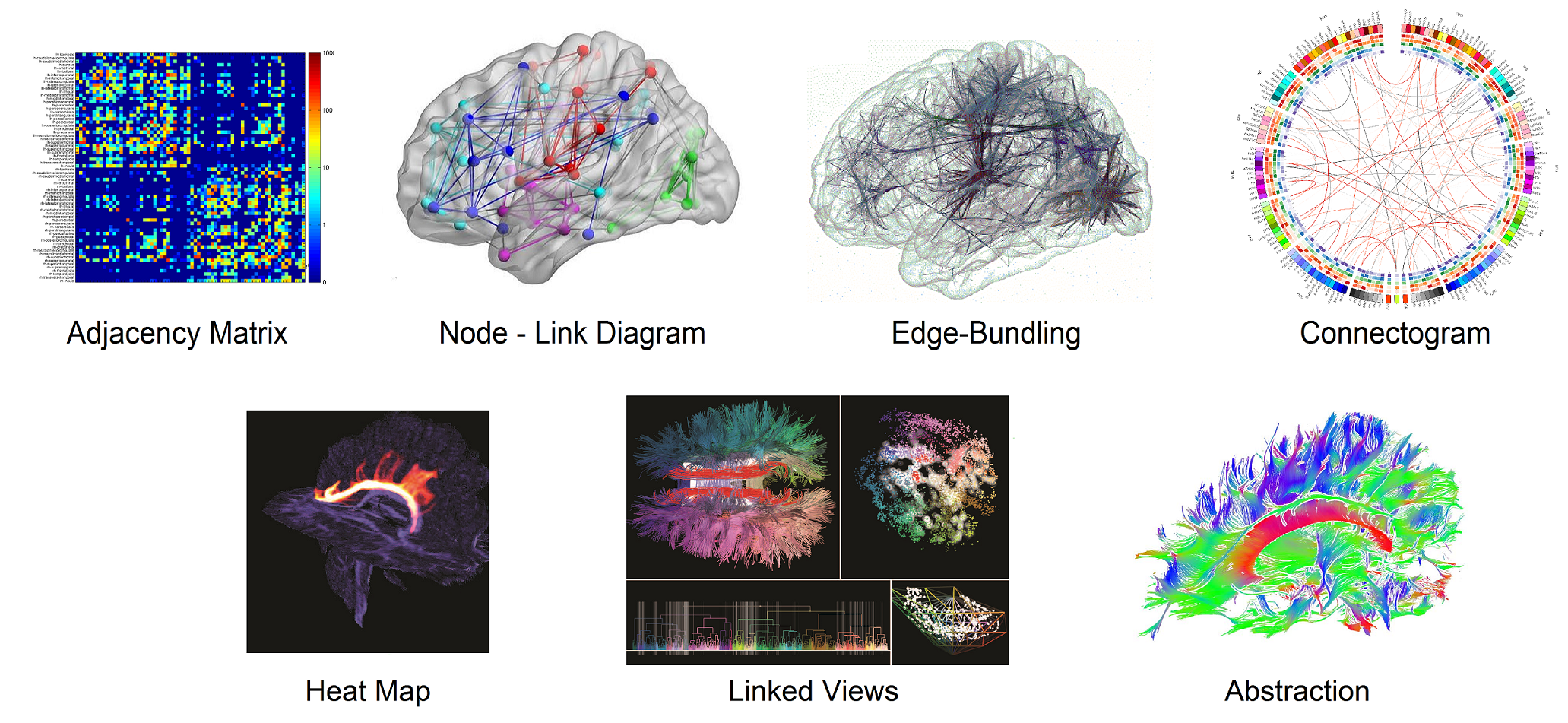}
 \caption{Different methods of visualization in connectomics. Top row, left to right: connectome: adjacency matrix (self prepared), node-link \cite{xia2013brainnet}, edge-bundling \cite{bottger2014three}, connectogram \cite{irimia2012circular}. Bottom row, left to right: tractography: heatmap \cite{mcgraw2007stochastic}, linked views \cite{jianu2009exploring}, abstraction \cite{everts2015exploration}.}
 \label{fig:all}
\end{figure*}

Over the past decade, the study of the human brain has progressed through advancements in the magnetic resonance imaging (MRI) and other neuroimaging technology. Those advancements have allowed neuroscientists to non-invasively probe the brain's structural and functional inter-regional connectivity and derive the human brain connectome \cite{sporns2005human}. High resolution MRI scans with submillimeter voxel size coupled with advanced non-linear registration algorithms allows the creation of brain label maps \cite{klein2009evaluation}. Those maps are created by registering brain MRI scans with a pre-segmented atlas by a highly experienced neuroscientist. Diffusion Weighted Imaging (DWI), also called diffusion MRI (dMRI), allows us to reconstruct the brain's white matter fiber tracts through a post-processing procedure called tractography. Counting the fibers interconnecting each pair of regions derived from the label map generates a brain structural connectome. Functional MRI (fMRI) measures the blood-oxygen-level dependent (BOLD) signal which represents the activation level of the different brain regions due to the execution of specific tasks. The functional connectome is then generated by computing the correlation coefficient of each pair of BOLD signal at the different brain regions. 

Connectomes are modeled as graphs which allows researchers to use graph theory mathematics to analyze them. Graph metrics can provide insights about the network \textit{topological} properties such as: its \textit{functional integration}, which is the network ability to combine information from its various parts; the \textit{clustering} or \textit{segregation} properties, which quantify the existence of groups (clusters or modules); and the network \textit{small-world} properties, which describe the balance between the functional integration and local clustering. A more extensive review of different graph metrics used in the field of connectomics can be found in \cite{rubinov2010complex}. Several important characteristics were derived for the healthy brain network such as small-worldness \cite{salvador2005neurophysiological, achard2006resilient}, clustering and modularity \cite{meunier2010modular}, and rich-club configuration \cite{van2011rich}. The functional connectome enables neuroscientists to determine the default mode network (DMN), or default state network, and was found to be the active brain interacting regions when the subject is at wakeful rest and not involved in a task \cite{buckner2008dmn}.

Group studies are used to study the effects of the onset of neurological and neuropsychological diseases, as well as age or injury on the human brain, and how it alters the connectome. In a group study, connectomes are generated for both healthy and disease populations. Graph metrics are then computed for the resultant networks, and then statistically significant variations in the metrics values can be determined. For a specific disease, researchers will often form a hypothesis based on previous findings in the literature and then attempt to correlate it to the alterations found in the collected graph data metrics \cite{angulo2016multi}. Although the use of graph metrics can provide a detailed aspect of the brain network, it does not deliver potentially useful spatial information, which could help the researcher to contextualize and interpret the results of a graph analysis. This mandates an efficient visualization for the connectome under study in order to help in the interpretation and diagnosis \cite{pfister2014visualization}. Moreover, visualization that supports effective comparison becomes crucial in group studies in which alterations in the disease group are more easily understood when analyzed in relation to healthy subjects. The contributions in this paper are as follow: 

\begin{itemize}[noitemsep,leftmargin=*]
\item We delineate visual analytics tasks for clinical neuroscientists, especially as related to group studies. 
\item We introduce a taxonomy of these tasks, providing a comprehensive survey of existing visualization software that supports connectome analysis.
\item We present \textit{NeuroCave}, a novel, web-based, immersive visual analytics system that facilitates the visual inspection of structural and functional connectome datasets. With \textit{NeuroCave}, brain researchers can interact with the connectome in any coordinate system or topological space, as well as group brain regions into different modules on demand. A default side-by-side layout enables simultaneous, synchronized manipulation in 3D that facilitates comparison tasks across different subjects or diagnostic groups, or longitudinally within the same subject.
\item We provide the opportunity for researchers to engage in immersive connectome analytics sessions while wearing portable VR headsets (e.g., Oculus Rift) or on stereoscopic displays (e.g., CAVE systems or 3D video walls).
\item We reduce visual clutter inherent in dense networks (both in 2D and in 3D views), by introducing a real-time, hardware accelerated edge bundling algorithm, and combining it with a user need-based edge selection strategy.
\item We present two real-world use cases that demonstrate how our visualization system can be used to identify patterns and support comparison tasks in order to understand differences between connectome datasets.
\end{itemize}

\section{Related Work} \label{review}

The term \textit{connectomics} was first coined by Sporns et al. \cite{sporns2005human} to describe the “wiring diagram” of the anatomical connectivity of the human brain. The interconnectivity found between parcellated brain regions as described above is considered the macroscale connectome. Meso- and micro- scale connectomes are at the local neuronal circuits and single neuronal cells levels respectively, and they currently require an invasive high resolution scans of dissected brains using microtome machinery. Friston defined functional connectivity as “the temporal correlation between spatially remote neurophysiological events” \cite{friston1994functional}. This paper focuses on visualization tasks relevant to macroscale functional and structural connectomes, derived from functional and diffusion-weighted MRI respectively.

\subsection{Visualization in connectomics} Initial efforts into visualizing connectomics data have focused on tractography data. The outcome of tractography computation is a set of lines (fibers) spanning the brain white matter area representing the axonal tracts of neuron cells. Volume rendering is used to overlay streamlines on top of the anatomical brain image \cite{sherbondyexploration, zhang2003visualizing, da2001visualizing, petrovic2007visualizing, schultz2007topological}. Often, a color code is used to identify the direction of the plotted fibers (see \autoref{fig:all}). The main task of such visualizations is to allow the user to select and explore specific fiber bundles such as the Corpus Callosum interconnecting the left and right hemispheres. Due to the huge number of generated fibers, on the order of $10^5$, a range of techniques are used to facilitate the navigation of the different fiber bundles. Some techniques make use of 2D representations of predefined fiber tracts that are synchronized together with a more conventional 3D visualization, such as 2D hierarchical tree-like grapha \cite{jianu2009exploring}, low-dimensional 2D embedding representations \cite{jianu2012exploring}, and 2D coronal, sagittal and axial projections of the fiber bundles \cite{chen2009novel} (see \autoref{fig:all}). Three dimensional visual abstraction of the fiber bundles is used in the recent work of Everts et al. \cite{everts2015exploration} (see \autoref{fig:all}). Heatmaps are also used in some techniques, where they are, for example, overlaid on top of anatomical isosurfaces \cite{mcgraw2007stochastic} (see \autoref{fig:all}). Segmented neurons and axons in mesoscale connectomes are also visualized using streamlines \cite{beyer2013connectomeexplorer, haehn2014design, ai2016neuroblocks}.

Two dimensional adjacency matrices can provide a good overview representation for large connectome datasets (see \autoref{fig:all}). Recent work enhances the readability and flexibility of the adjacency matrix \cite{alper2013weighted, ma2015visualizing}, but have not been incorporated into software platforms that are readily available for neuroscientists. Moreover, the use of adjacency matrices can hinder users in performing some visual analysis tasks, such as detecting graph alterations in group studies \cite{ghoniem2005readability, keller2006matrices}. The most widely used representation for connectome visualization are 3D node-link diagrams, in which nodes are positioned relative to their corresponding anatomical locations, and in which links represent the structural or functional connectivity between the nodes. Many examples of visualization applications are presented in \autoref{tab:table1}. An inherent problem of node-link diagrams is the difficulty of effectively displaying the total available links in dense networks. In dense networks, visual clutter due to a preponderance of edge crossings can adversely affect the accuracy and completion time for \textit{trend} tasks (assessing change in edge weight of a node's connections), \textit{connectivity} tasks (assessing the connectivity of common neighbors) and \textit{region identification} tasks (identifying the region with the most changes) \cite{alper2013weighted}. One technique to reduce such clutter is to combine visually compatible edges into bundles according to a compatibility metric, hence the name \textit{edge bundling}, first introduced by Holten in \cite{holten2006hierarchical}. Recently, some connectome visualization projects have utilized edge bundling. Two dimensional edge bundling is used by McGraw \cite{mcgraw2015graph} and Yang et al. \cite{yang2017blockwise}, while 3D edge bundling is used in \cite{bottger2014three, bottger2014connexel} for representing functional connectivity that shows high levels of common interconnections (see \autoref{fig:all}). \textit{NeuroCave} also utilizes edge bundling in order to reduce visual clutter in large connectome datasets that may have 2500 or more interconnected nodes (see \autoref{fig:case2}).

The connectogram \cite{irimia2012circular}, a recent connectome visualization technique, is a node-link diagram in a circular layout (see \autoref{fig:case2}). Names of brain regions are positioned along the perimeter of a circle, which is split into two halves according to their hemispheric affiliation. Each hemisphere is further divided into the different brain lobes. The inner sphere contains multiple colored rings, each representing a specific metric. The regions are interconnected within the circle using curved lines. This technique prevents some of the clutter that is found in other network visualizations. However, it is harder to correlate anatomical structures with connectivity using a circular diagram like the connectogram, as the multiple rings may make it difficult for a user to make sense of the data \cite{burch2014benefits}. \autoref{tab:table1} contains a list of different visualization tools in the field of connectomics and their capabilities. Although previous efforts have surveyed visualization in connectomics, such as Margulies et al.~\cite{margulies2013visualizing} and Pfister et al.~\cite{pfister2014visualization}, our survey is the first to comprehensively explore recent connectomics visualization software in terms of their ability to support group studies analysis tasks. Moreover, we catalog these software tools by the primary connectome dataset types they support.

\subsection{Connectome Analysis Tasks}
\label{ss:CAT}
Extending the user task definitions presented in Alper et al.~\cite{alper2013weighted} through a thorough investigation of connectome visualization projects and surveying neuroscientists that work with group studies, we have identified visual analytics tasks for clinical neuroscientists:

\begin{enumerate}[noitemsep,leftmargin=*]
\item [\textbf{T1}] Identify regions responsible for specific cognitive functions and study their interactions with other regions.
\item [\textbf{T2}] Compare individual networks to the mean or group average connectome. In group studies, individual variations as well as joint network characteristics are studied in order to identify commonalities or differences.
\item [\textbf{T3}] Identify the effect of structural connectivity on the functional activity of the brain. Comparing both structural and functional at the same time to reveal the complex mappings between them \cite{bullmore2009complex, honey2007network, honey2009predicting}.
\item [\textbf{T4}] Identify individual or group changes occurring on the structural or functional connectivities due to the onset of disease \cite{sanz2010loss} or aging \cite{dosenbach2010prediction} as well as gender differences. Moreover, researchers can assess its restoration in drug studies \cite{tadayonnejad2016pharmacological}.
\item [\textbf{T5}] Identify dynamic changes in structural and functional connectivity over time for both within subject and between subjects \cite{crossley2016connectomic}.
\item [\textbf{T6}] Identify structural re-routing occurring after a brain injury or damage with rehabilitation training and its effect on functional connectivity. Similarly, in case of neurosurgery, it is important to identify affected structural pathways and predict the corresponding loss in motor and cognitive abilities after the procedure.
\end{enumerate}

Although each of the the visualization software tools listed in \autoref{tab:table1} may partially address many of these tasks, none provides a visualization that can directly facilitate T2--T5, involving various types of comparison between datasets, since they all lack the ability to simultaneously load and synchronize a comparative visualization of multiple connectomes. The user needs to open multiple instances of the application and usually requires multiple monitors in order to visually be able to compare the structural and functional connectomes of the same subject or two subjects belonging to different groups. Clearly, all user actions will not be synchronized which makes it even harder for assessing visual differences. Applications implemented in scripting languages such as R and MATLAB provide the user with the flexibility to customize views. However, this requires additional efforts as well as programming expertise. Introducing a novel side-by-side layout, \textit{NeuroCave} helps neuroscientists and researchers efficiently execute T1--T5, which involve comparative analyses, and to simultaneously spot changes occurring within and across subjects. It is important to state that \textit{NeuroCave} does not target tractography-related usages, such as the sixth task (T6), which, although an important area of connectomics visualization, is usually a requirement by neurosurgeons rather than clinical neuroscientists (who are the intended audience for our visualization system). 

\subsection{Web-based VR} While most commonly used visualization tools are dedicated desktop applications, web-based implementations, such as \textit{Slice:Drop}~\cite{haehn2013slice} or \textit{BrainBrowser}~\cite{sherif2015brainbrowser}, free the user from being attached to a specific operating system~\cite{pieloth2013online}. \textit{NeuroCave} is also a web-based application and runs in any modern browser. The use of stereoscopic techniques can provide a more immersive way to explore brain imaging data~\cite{rojas2014stereoscopic}, H{\"a}nel et al. find that healthcare professionals perceive the increased dimensionality provided by stereoscopy as beneficial for understanding depth in the displayed scenery \cite{hanel2014interactive}. Moreover, Ware and Mitchell find that the use of stereographic visualizations reduces the perception error rate in graph perception for large graphs with more than 500 nodes~\cite{ware2008visualizing}, and Alper et al.~\cite{alper2011stereoscopic} observed that, when coupled with highlighting, stereoscopic representations of 3D graphs outperformed their non-immersive counterpart. \textit{NeuroCave} lets the user move between desktop and VR environments for interactively exploring 3D connectomes.

\begin{table*}[!htbp]
% \begin{landscape}
% \thispagestyle{empty}
\caption{Neuroimaging connectomic software. This table indicates the visual analysis tasks (Section~\ref{ss:CAT}) supported by each of these software tools.}
\label{tab:table1}
\noindent
\centering
\includegraphics[width=\textwidth]{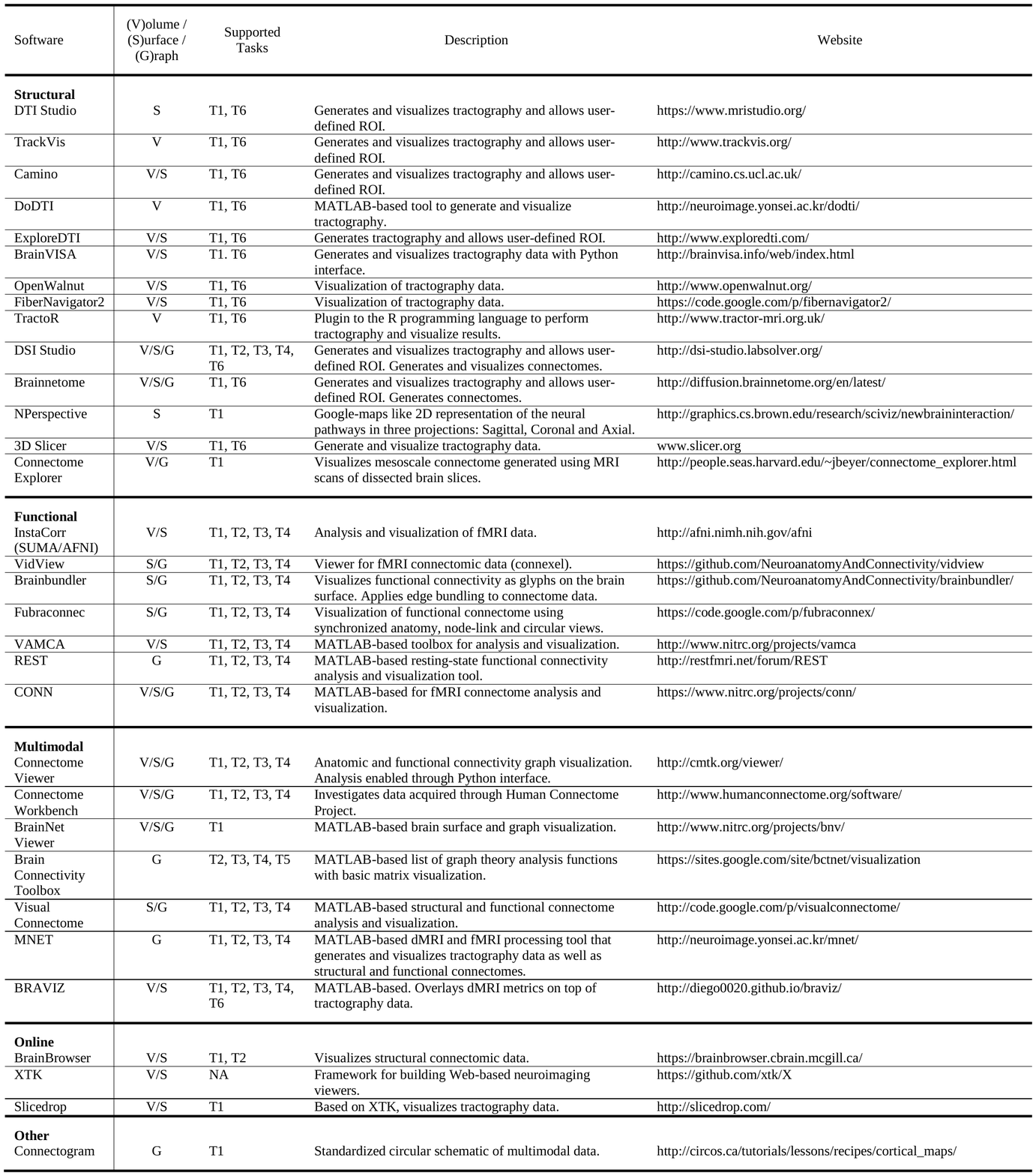}
% \end{landscape}
\end{table*}

\subsection{Connectome Topology} The techniques of linear and nonlinear dimensionality reduction have been widely used in the field of brain imaging, specially for processing fMRI data \cite{thirion2004nonlinear, calhoun2009review}. Recently, nonlinear dimensionality reduction techniques, such as isomap and t-distributed stochastic neighbor embedding (t-SNE), have been applied onto the human connectome in order to find the brain's intrinsic geometry \cite{allen2015intrinsic}. Using such embedding, new 3D topologies for the brain were found other than the regular anatomical one. Such topologies depends on the internodal graph shortest path length computed using Dijkstra's algorithm \cite{dijkstra1959note}. Nodes possessing efficient connectivity to the rest of the brain were found to be positioned near the origin of those new topologies inferring their importance. NeuroCave facilitates finding topological alterations in group studies. This helps researchers in studying the connectome topology of healthy subjects, such as the DMN, and comparing it with diseased groups. In DMN, important hub nodes, such as the precuneus (see \autoref{fig:case2}), possess a strong connectivity with the rest of the brain and are located near the center of the brain's intrinsic geometry \cite{allen2015intrinsic}. A deviation from such topology in disease groups need to be studied for illnesses associated with DMN alterations such as Alzheimer's disease which is known to disrupt it \cite{buckner2008dmn}.

\section{The \textit{NeuroCave} Software System} \label{neurocave}

\begin{figure}[tb]
 \centering 
 \includegraphics[width=\columnwidth]{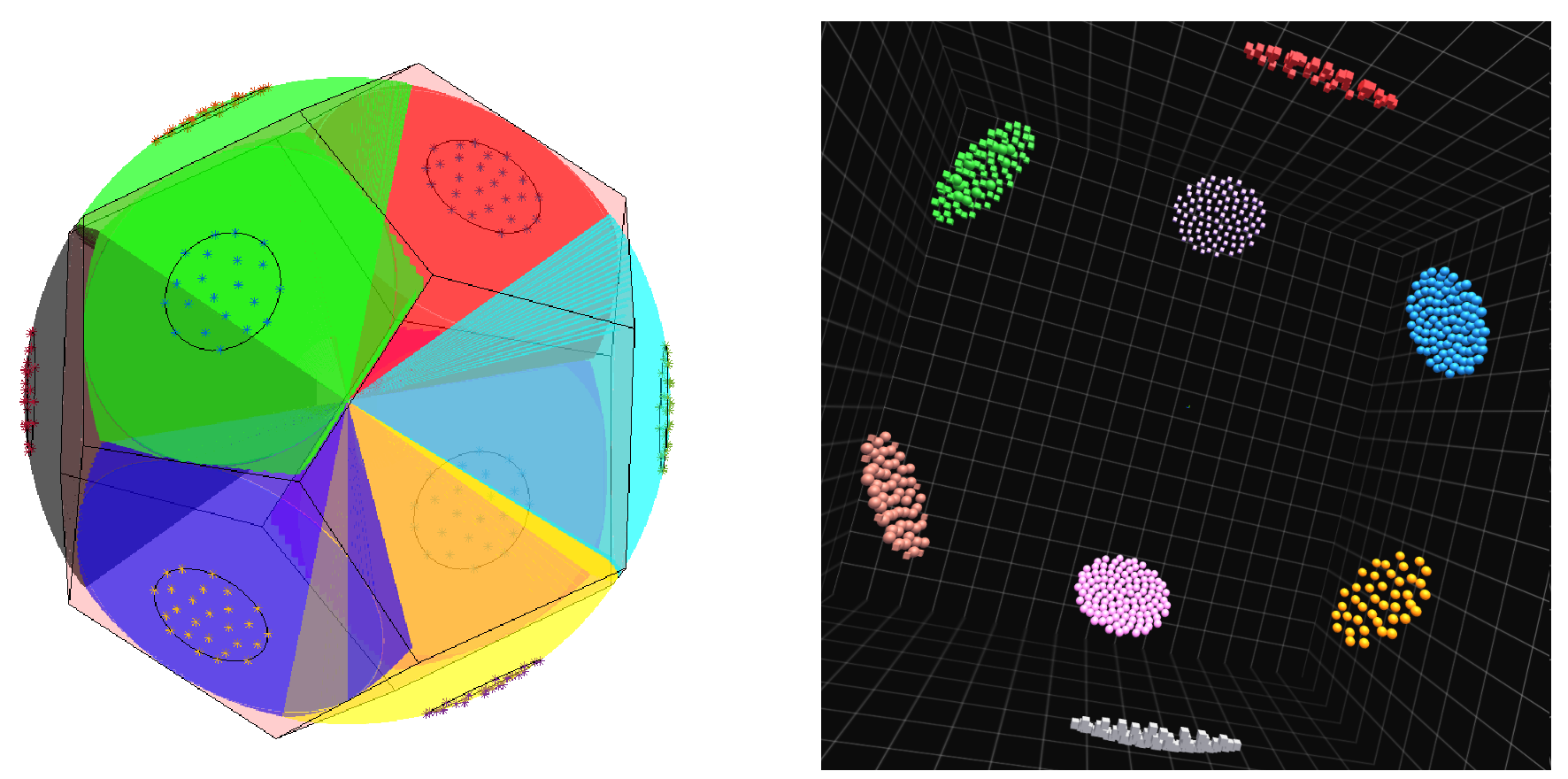}
 \caption{The figure shows the proposed method of distributing clustered nodes. The nodes are distributed on half the spherical cap covering a platonic solid face. The chosen platonic solid must have a number of faces that is equal to or larger than the number of clusters (for 8 clusters a dodecahedron was chosen).}
 \label{fig:platonics}
\end{figure}

\begin{figure*}[!htbp]
 \centering
 \includegraphics[width=\textwidth]{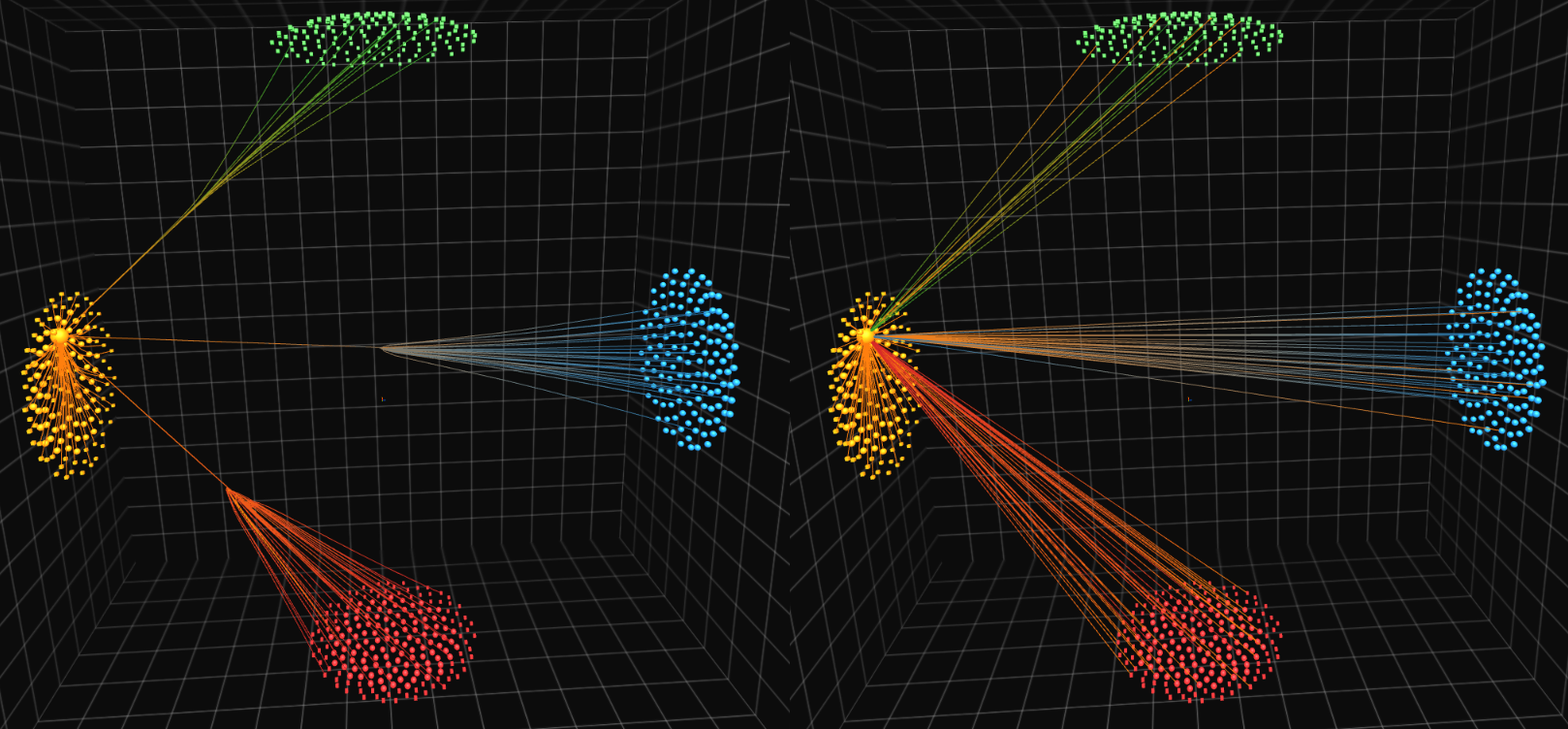}
 \caption{Edge bundling (left) versus no edge bundling (right) in clustering space displaying four clusters. Notice the clutter caused by the edges in the right panel versus the edge bundling case. Also, notice the edge color gradient that depends on the nodes color and their relative strengths: the orange sphere possesses a greater strength compared to the blue nodes it is connected to, while it possesses a lesser strength compared to the green nodes it is connected to. \textit{NeuroCave} enables users to toggle between edge bundling and edge coloring modes on demand.}
 \label{fig:eb}
\end{figure*}

\textit{NeuroCave} is implemented as a web-based application. The Javascript WebGL-based graphics library \textit{three.js} \footnote{\url{http://threejs.org}.} was used for the 3D rendering and visualization functionalities. It runs on all major web browsers, and is thus platform independent. The default view is formed of two side-by-side 3D rendering views (see \autoref{fig:teaser}). Each view allows the interactive visualization of a connectome as a node-link diagram.\\
\textbf{Group Visualization.} The application loads subjects data from a study folder. The folder should contains all adjacency matrices as well as the corresponding topological and clustering information of the subjects within the study. An indexing file states the subject ID and its corresponding data files. Each study requires a predefined Atlas that provides numerical labels and their corresponding anatomical names to each node. \textit{NeuroCave} directly supports three Atlases: FSL-based which consists of 82 labels of FreeSurfer \footnote{\url{https://surfer.nmr.mgh.harvard.edu/}.}, brain hierarchical Atlas (BHA) made of 2514 labels \cite{diez2015novel} and Harvard-Oxford Atlas made of 177 labels.\footnote{\url{http://neuro.imm.dtu.dk/wiki/Harvard-Oxford_Atlas}.} However, additional Atlases can be easily created following the pre-existing ones as example.\\
\textbf{Topology Visualization.} The application layouts the nodes according to the provided topological information. Topologies can be the anatomical positioning or an applied transformation in some abstract space as explained in Section \ref{review}. The available topologies are automatically identified by the application.\\
\textbf{Clustering Visualization.} Clustering information is input as a vector of integer values. Each value represents a different module or cluster. Since there are no prespecified positions for clusters, we exploit the geometrical properties of platonic solids. In brief, a platonic solid is a regular, convex polyhedron constructed by congruent regular polygonal faces with the same number of faces meeting at each vertex. Five platonic solids exist: tetrahedron, cube, octahedron, dodecahedron and icosahedron, with four, six, eight, twelve, and twenty faces, respectively. Based on how many clusters are generated, a suitable platonic solid is chosen such that its number of faces is greater than the number of these clusters. The glyphs of each cluster are then equally distributed (according to the sunflower algorithm~\cite{vogel1979better}) over half of the spherical cap covering the corresponding face of of a platonic solid embedded in a sphere (see \autoref{fig:platonics}). This enables the user to ``enter'' into the geometry (i.e. into the ``NeuroCave'') via one of the unpopulated face(s), providing a more immersive experience of the data.\\
\textbf{Node visualization.} We utilize two different glyphs (spheres and cubes) to differentiate between left and right hemisphere affiliation. Nodes can be colored according to lobar or modular information. Controlling nodal transparency is also possible according to their color scheme. Three modes exist: opaque, semi-transparent or totally transparent (invisible). Glyphs size is also user adjustable.\\
\textbf{Edge Visualization.} As mentioned in Section~\ref{review}, edge clutter can be a problem in the visualization of dense node-link diagrams. Our approach to overcome this problem is based on two steps. First, we provide the option to hide all edges by default (i.e., to show only the nodes). The user can then select a root node, and all connected edges stemming from this node will be displayed. Second, to overcome the clutter occurring from edge crossings, we use the force directed edge bundling (FDEB) algorithm to group edges going in the same direction~\cite{holten2009force}. FDEB is an iterative algorithm that consists of cycles. In each cycle, we subdivide an edge into a specified number of points (we chose 6 cycles, and we double the number of points each cycle, ending up with 64 subdivision points plus the two original points of the edge). After the subdivision, we iteratively move each subdivision point in an update step to a new position determined by modeling the forces among the points. Each point is affected by the sum of spring and electrostatic forces ($F_s$ and $F_e$). For an edge \textit{i}, at a subdivision point $p_{ij}$, $F_s$ is defined as:

\begin{equation}
F_s = k_p(\| p_{i(j-1)}-p_{ij}\| + \| p_{ij} - p_{i(j+1)}\| )
\end{equation}
where $k_p$ is a spring constant and $p_{i(j-1)}$ and $p_{i(j+1)}$ are the neighboring points of $p_{ij}$, the $\|.\|$ is the length. The electrostatic force $F_e$ is defined as

\begin{equation}
F_e =  \sum_{m \in E} \frac{1}{\|p_{ij}-p_{mj}\|}
\end{equation}

\noindent where $E$ is a set of compatible edges with the current edge, $p_{mj}$ is the corresponding subdivision point to point $p_{ij}$ in the edges belonging to the set $E$. The compatibility metrics used to define the set $E$ are defined in details in Holten et al. \cite{holten2009force}. Each cycle contains a prespecified number of update steps. 

Our original Javascript implementation of \textit{NeuroCave} turned out to be too slow for large numbers of edges (more than 500), preventing a real-time experience. Therefore, we used (and enhanced) a WebGL texture-based implementation suggested by Wu et al.~\cite{wu2015texture}. The FDEB algorithm is parallelizable since the division and update operations are performed on each point independently. The texture-based method stores the subdivision points in a 2D GPU texture, where each row represents the 3D coordinates of points belonging to the same edge. Since write operations are unavailable to GPU textures in WebGL, a ping-pong algorithm \footnote{\url{https://www.khronos.org/opengl/wiki/Memory_Model}.} is used to write results to a framebuffer object (FBO). Two shaders are utilized: the first performs the subdivision operation, and the second executes the update steps. In Wu et al.'s implementation, the limitation on the texture size limits is not addressed. GPU textures possess a limitation on their sizes. Hence, a large number of edges can not be fit in one texture. We enhanced the algorithm using tiling. Since the total number of points of each edge after all cycles will be known ahead (64 subdivision + 2 end points), we tile the edges when the maximum number of possible rows per texture is achieved. Our texture-based implementation can bundle the closet 1000 edges to the selected node at interactive rates on a desktop computer (Intel Core i7, 3.4 GHz CPU, Nvidia GTX 1070 GPU card and 32GB RAM).

Each edge is colored according to a two color gradient chosen according to the interconnected nodes colors. The gradient is skewed towards the node possessing the higher nodal strength (the sum of weights of links connected to the node) value (see \autoref{fig:eb}). For an edge made of $n$ points the color, $C_{out}$, at point $i$ is 

\begin{equation}
C_{out} = C_2 + (C_1-C_2)R
\end{equation}

\begin{equation}
R = ( r^2(p_2-0.5) + r(0.5-p_2^2) )/P_1P_2
\end{equation}
where $C_1$ and $C_2$ are the nodes, $r = i/(n-1)$, $P_1 = S_1/S_1+S_2)$, $P_2 = 1-P_1$, $S_1$ and $S_2$ are the nodes strengths. This allows the user to recognize the strength of the selected node with respect to its interconnected neighbors which helps identifying strong and important nodes and hubs as well as highlight the reason for modular changes when occurred in group studies as will be shown in Section \ref{results}.\\
\textbf{Virtual Reality.} In addition to the standard 3D manipulations of panning, rotating, and zooming, NeuroCave supports the Oculus Rift and the Oculus Rift Touch controllers. The Touch controllers are a pair of VR input devices that track each hand, enabling an effective gesture-based manipulation of the VR environment. The user selects the preview area to be explored in VR and then uses the thumbsticks of the Touch devices to navigate the visualized connectome through panning, rotating, and zooming. Nodal selection is enabled via a two step procedure: first, pressing the grip button lets the user point at and highlight a node; second, pressing the index button selects the highlighted node.

\begin{figure*}[p]
 \centering
 \includegraphics[width=\textwidth]{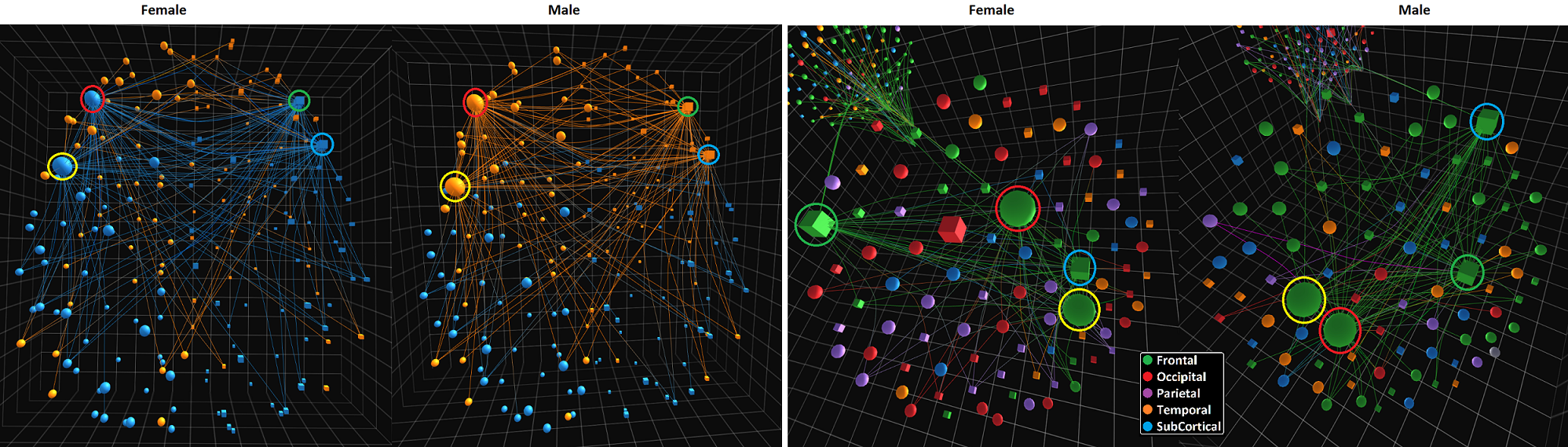}
 \caption{Left and right frontal poles (red and green rings encircling the sphere and cube glyphs), left precentral gyrus (yellow ring) and the right inferior frontal gyrus (blue ring) nodes selected in anatomical (left panel) and clustering (right panel) spaces for female and male average connectomes. The color code represents the two clusters in the anatomical space, while in the clustering space, it represents the lobe affiliation. Notice the change in group affiliation of the four nodes between the female and male connectomes (left panel).}
 \label{fig:caseaa}
\end{figure*}

\begin{figure*}[p]
 \centering
 \includegraphics[width=\textwidth]{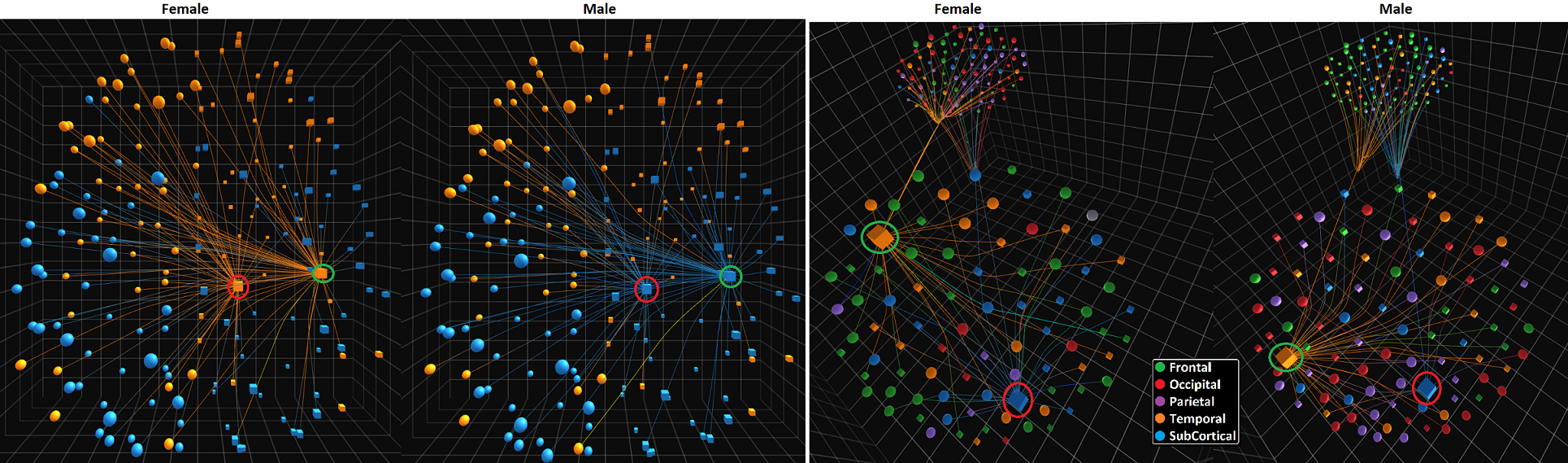}
 \caption{Right hippocampus (red ring) and superior temporal gyrus (green ring)  nodes selected in anatomical (left panel) and clustering (right panel) spaces for female and male average connectomes. Color code is similar to the above figure. Notice the change in group affiliation of the four nodes between the female and male connectomes (left panel).}
 \label{fig:caseab}
\end{figure*}

\begin{figure*}[p]
 \centering
 \includegraphics[width=\textwidth]{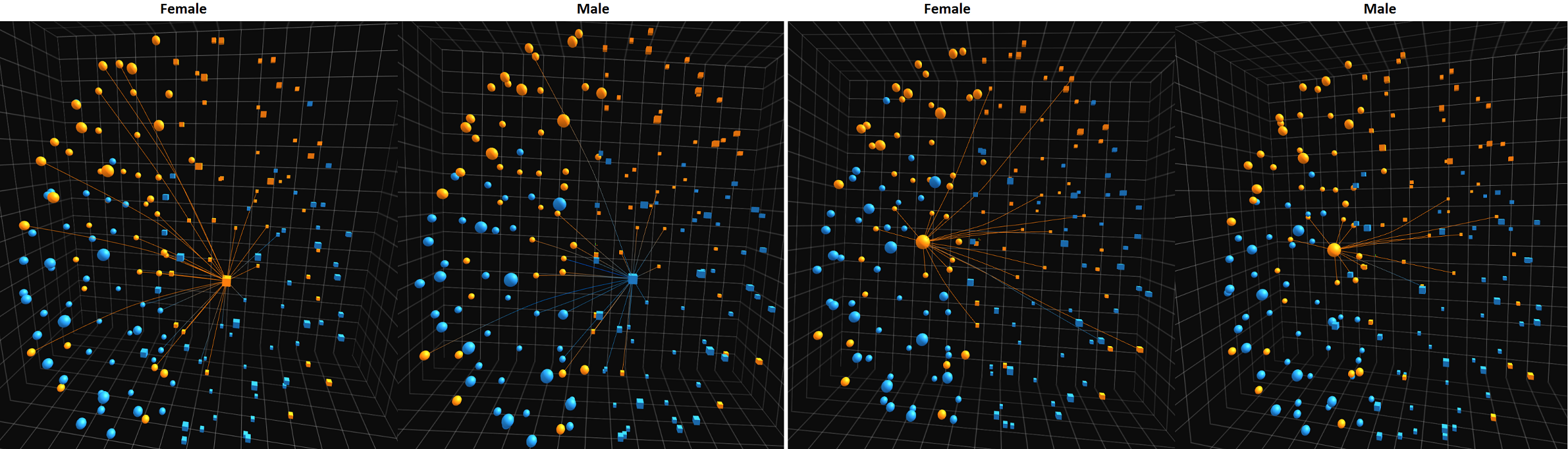}
 \caption{Right hippocampus (left panel) and left hippocampus (right panel) nodes selected in the anatomical space for female and male average connectomes. Note the tendency in females for both left and right hippocampus to be functionally connected to the contralateral frontal lobe (but not in males). Note that such differences are not previously known to our clinicians and neuroscientists. Here, the color code represents connectome's hierarchical modularity, represented as a dendrogram, at the most global level (2 modules).}
 \label{fig:caseac}
\end{figure*}

\begin{figure*}
 \centering
 \includegraphics[width=\textwidth]{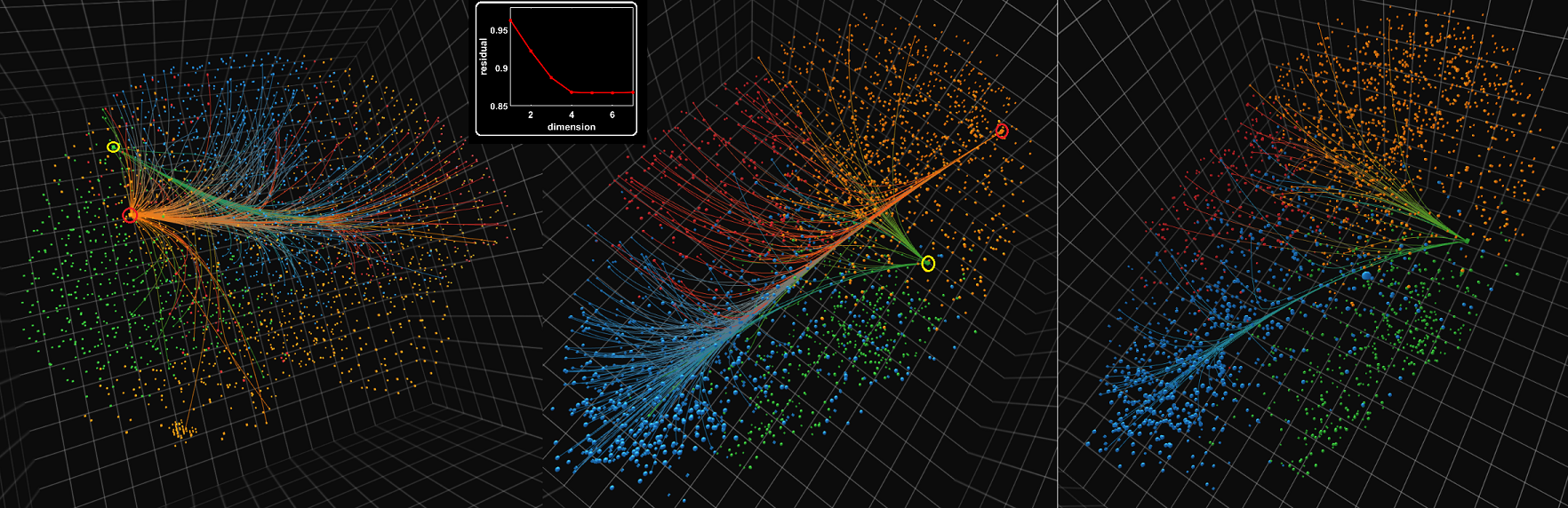}
 \caption{Connectivity emerging from the anterior (red ring) and posterior (yellow ring) parts of the precuneus in anatomical (left panel) and isomap ``intrinsic'' (middle panel) spaces. Right panel: Connectivity emerging from the posterior part of the precuneus visualized in the intrinsic space. The color code represents the modular structure of the connectome consisting of 4 modules. Note that the orange community contains the default mode network. The inset plot shows the residual geodesics for the first 10 dimensions of the isomap dimensionality reduction algorithm.}
 \label{fig:case2}
\end{figure*}

\section{Results and Discussion} \label{results}
In this section, we present two use cases that demonstrate how clinical neuroscientists can perform the tasks described in Section~\ref{review}.

\subsection{Use Case 1}
Our first use case investigates the sex-specific resting-state functional connectomes in the F1000 repository, a large 986 subjects publicly-available resting-state fMRI connectome dataset.\footnote{\url{https://www.nitrc.org/projects/fcon_1000/}.} The following post-processing was performed: first, to eliminate the potential confounding effect of age, we only included subjects between 20 to 30 years old (319 females at 23.25 $\pm$ 2.26 years of age and 233 males at 23.19 $\pm$ 2.35). We then constructed hierarchical modularity in the form of 4-level dendrograms (16 clusters at the most local level)~\cite{zhanimportance}. Previously, rigorous permutation testing established sex differences already statistically significant at the first level (2 clusters; p = 0.0378) for the following six regions: posterior part of left and right frontal pole (near the fronto-temporal junction), left precentral gyrus, right hippocampus, right superior temporal gyrus posterior division (sTG-p) and the right inferior frontal gyrus pars opercularis (POP) (for more detail about the statistical analysis, refer to \cite{zhanimportance}).

In order to interpret these sex differences, male and female group-averaged connectomes were visualized in NeuroCave. We highlighted the connectivity of the aforementioned six regions in both the anatomical and clustering (1st hierarchical level) spaces and applied a threshold to restrict edges to values with absolute correlation values for fMRI BOLD signals larger than 0.1  (see \autoref{fig:caseaa}, \autoref{fig:caseab}). From the clustering space, affiliation patterns for the right hippocampus and sTG-p differ in that in women they are clustered with other frontal and temporal ROIs as part of the default mode network (DMN), while in males they are clustered with other parietal and occipital non-DMN ROIs. Notably, the affiliation differences are opposite for the left and right frontal poles, the left precentral gyrus and the right POP, such that they are more clustered along with several non-DMN parietal and occipital ROIs in the average female connectome (see \autoref{fig:caseaa}). 

The left and right fronto-temporal junctions are part of the larger language system, with the right POP (functionally coupled with its homologous area on the left that forms the Broca's language area) linked to the processing of semantic information \cite{heim2005role}, and the superior temporal gyrus involved in the comprehension of language (as well as in the perception of emotions in facial stimuli \cite{bigler2007superior}). Thus, the observed connectivity differences are related to well-known sex differences in language and emotion/affect processing, as well as differences in self-referential/autobiographical information retrieval.  By contrast, the hippocampus is known to play an important role in the formation of new memory, retrieval of declarative long-term memory, and the management and processing of spatial and spatiotemporal working memory. The modular affinity between the right hippocampus and other non-DMN regions in the parietal and occipital lobes in males may thus be related to their established advantage in spatial tasks, including spatial visualization, perception and mental rotation \cite{linn1985emergence} (visual system is heavily composed of the occipital lobe responsible for first-level visual processing, while part of the parietal lobe is instrumental for visuospatial skills).

Lastly, closer inspections of the hippocampal functional connectivity between women and men further revealed differences that are not known previously to our clinicians (Fig 7). Indeed, in the female connectome there is a strong tendency for both left and right hippocampus to functionally communicate with the contralateral frontal lobe, a phenomenon that is not visually present in the male connectome. The discovery of such subtle differences is dependent upon an iteratively explorative visualization process, only possible with the comprehensive suite of tools implemented in NeuroCave.

This use case demonstrates the effectiveness of NeuroCave is supporting tasks T3 and T4, enabling  neuroscientists to better understand neurological gender differences in connectome datasets and to   observe how these differences relate to various psychological studies.

\subsection{Use Case 2}
Our second use case explores a resting-state fMRI high-resolution dataset consisting of 2514 regions-of-interest publicly available at NITRC.\footnote{\url{http://www.nitrc.org/frs/?group_id=964}.} Most functional connectome studies treat the negative correlation entries (anticorrelated BOLD signals) in the network by either taking the absolute value or clamping to zero (heuristically chosen) which affects all network metrics' computations. Instead, we applied a recently proposed probabilistic framework which more rigorously accounts for negative edges \cite{zhanimportance}. For an $N \times N$ functional connectome, the framework estimates the probability that an edge,  $e_{ij}$, is positive or negative using the connectomes of a group of subjects ($i$ and $j$ varies from 1 to $N$). The edge positivity $EP_{ij}$ and edge negativity $EN_{ij}$ form a complementary pair, since $EP_{ij} + EN_{ij} = 1$, and thus can be jointly coded using the angle of a unit-length vector: $\theta_{ij} = \arctan ( \sqrt{EN_{ij}/EP_{ij}} )$  which varies from 0 to 90 degrees. Using \textit{dissimilarity graph embedding}, each node, $i$, is then embedded in an n-dimensional space at the coordinates $(\theta_{i1}, \theta_{i2}, ... \theta_{iN})^T$. In order to reduce the dimensionality of the resultant topology, we applied the nonlinear dimensionality reduction isomap algorithm \cite{tenenbaum2000global} that aims to preserve geodesic distances in a lower-dimensional space (i.e., the ``\textit{intrinsic space}''). Separately, we determined the community structure by maximizing the Q-modularity metric \cite{newman2006modularity}.

From the inset plot in \autoref{fig:case2}, it is clear that the degree of isometric embedding levels off after the fifth dimension, thus suggesting that the intrinsic topology of the resting-state functional connectome has a dimension of five, a novel finding that merits further research. To enable 3D visualization of the transformed topology, we retained the first three dimensions of isomap and visualized the modular structure of the brain in both the anatomical space as well as this novel intrinsic topological space (Fig 8). As expected, the nodes assigned to the same community are positioned close to one another in this novel topology. To illustrate how neuroscientists explore this complex topological space and gain further insight into the brain, we selected two nodes that belong to the anterior and posterior part of the precuneus. Although the nodes are anatomically close to each other, they are known to be functionally distinct (thus belong to different modules). Indeed, the anterior part of the precuneus is an important region of the DMN known to be responsible for self-referential imagery (thinking about self) and is involved in autobiographical tasks and self-consciousness, thus activated during ``\textit{resting consciousness}''~ \cite{cavanna2007precuneus}. As we can see in Fig. 8, in this intrinsic space the anterior part of the precuneus, while assigned to the orange module (DMN), exhibits diverse connections with various regions of the brain: the sensori-motor module (blue) and the frontoparietal executive or task-positive system (red). By contrast, the posterior precuneus is part of the visual system (green) and has a relatively restricted pattern (compared with its more anterior counterpart) of connectivity with the rest of the brain. Notably, such connectivity differences only become visually apparent when neuroscientists visualize in this novel space.

This use case highlights how our visualization system  is effective at supporting tasks T1 and T2, enabling  neuroscientists to explore high density connectomics data comprising a large number of ROIs in order to identify and further understand the specific functionality of different brain regions. Moreover, the side-by-side visualization enables users to reason about the relationships between the anatomical and the intrinsic topology, facilitating further insight into how the same brain region can take part in different tasks.

\section{Conclusion and Future Work}

In this paper, we presented \textit{NeuroCave}, a novel VR-compatible visualization environment for exploring and analyzing the human connectome. We  also performed a task taxonomy for neuroscientists and researchers in the field of connectomics. NeuroCave facilitates comparison tasks of two connectomes, an activity that clinical neuroscientists often require for the analysis of group studies. Our system includes various visualization enhancements, such as edge-bundling to reduce edge clutter and edge gradient coloring to help identify nodal strengths. Moreover, it enables users to explore connectomic data in an immersive VR environment, which can help to improve their perception of the data under study. The GPU is extensively used in NeuroCave for hardware acceleration of both computation and visualization tasks. We demonstrated the effectiveness of the system using two real-world use cases in which neuroscientists were able to use NeuroCave to effectively perform research and analysis tasks. 

However, many challenges in visualizing the brain connectome  remain. Future work will adapt our system to support temporally-varying dynamic connectome datasets (T5). We also will continue to work with clinical neuroscientists to integrate additional analytic tools to assist with diagnosis and treatment of patients. Finally, our side-by-side visualization framework can be extended to other domains that require group study analysis, such as cancer biology, where researchers are interested in comparing experiment vs. control networks or disease vs. normal gene networks in order to better understanding cancer and other disease genomes.

%% if specified like this the section will be committed in review mode
%\acknowledgments{
%The authors wish to thank A, B, C. This work was %supported in part by a grant from XYZ.}

%\bibliographystyle{abbrv}
\bibliographystyle{abbrv-doi}

\bibliography{neurocave}
\end{document}